\documentclass[a4paper,11pt]{article}
\pdfoutput=1
\usepackage{jheppub}

\usepackage{amssymb,amsmath,amsfonts}
\usepackage[normalem]{ulem}
\usepackage[utf8x]{inputenc}
\usepackage{slashed}
\usepackage{graphicx}
\usepackage{tabularx}
\usepackage{here}
\usepackage{color}
\usepackage{csquotes} 
\usepackage{comment}
\usepackage{mathrsfs}
\usepackage{float}
\usepackage{ascmac}
\usepackage{multirow}
\usepackage{longtable}
\usepackage{caption}
\usepackage{bm}
\usepackage{ulem}
\usepackage{tikz}
\usepackage{tikz-feynman}
\tikzfeynmanset{compat=1.1.0, warn luatex=false}
\usetikzlibrary{decorations.pathmorphing,positioning,shapes.geometric,calc,arrows.meta}

\usepackage{colortbl} 

\usepackage{cancel}

\usepackage[italicdiff]{physics}
\usepackage{wrapfig}

\makeatletter
\newcommand*\rel@kern[1]{\kern#1\dimexpr\macc@kerna}
\newcommand*\widebar[1]{%
  \begingroup
  \def\mathaccent##1##2{%
    \rel@kern{0.8}%
    \overline{\rel@kern{-0.8}\macc@nucleus\rel@kern{0.2}}%
    \rel@kern{-0.2}%
  }%
  \macc@depth\@ne
  \let\math@bgroup\@empty \let\math@egroup\macc@set@skewchar
  \mathsurround\z@ \frozen@everymath{\mathgroup\macc@group\relax}%
  \macc@set@skewchar\relax
  \let\mathaccentV\macc@nested@a
  \macc@nested@a\relax111{#1}%
  \endgroup
}
\makeatother

\numberwithin{equation}{section}

\preprint{
\begin{minipage}{5cm}
\small
\flushright
EPHOU-25-018\\
KYUSHU-HET-340
\end{minipage}}

\title{GUT-motivated non-invertible symmetry as a solution to the strong CP problem 
and the neutrino CP-violating phase}

\author{Tatsuo Kobayashi$^{1}$,} 
\author{Hajime Otsuka$^{2}$,}
\author{Morimitsu Tanimoto$^{3}$, \\ and} 
\author{Tsutomu T. Yanagida$^{4,5}$}
\affiliation{
$^1$Department of Physics, Hokkaido University, Sapporo 060-0810, Japan\\
$^2$Department of Physics, Kyushu University, 744 Motooka, Nishi-ku, Fukuoka 819-0395, Japan\\
$^3$Department of Physics, Niigata University, Ikarashi 2-8050, Niigata 950-2181, Japan\\
$^4$Kavli IPMU (WPI), UTIAS, University of Tokyo, Kashiwa, 277-8583, Japan\\
$^5$Tsung-Dao Lee Institute \& School of Physics and Astronomy, Shanghai Jiao Tong University, China
}
\emailAdd{kobayashi@particle.sci.hokudai.ac.jp}
\emailAdd{otsuka.hajime@phys.kyushu-u.ac.jp}
\emailAdd{morimitsutanimoto@yahoo.co.jp}
\emailAdd{tsutomu.tyanagida@sjtu.edu.cn}

\abstract{
The unsuppressed CP violation in QCD is a problem in the standard model. If we have some mechanism to guarantee real determinants of the quark mass matrices, the vanishing physical vacuum angle $\bar \theta$ indicates the CP invariance at the fundamental level. Thus,  the small ${\bar \theta}$ is technically natural, since we have an enhanced CP symmetry in the limit of the vanishing $\bar \theta =0$. In fact, it was proved that the vacuum angle is never renormalized up to the four-loop level once it is fixed at 0 value at some high energy scale. The purpose of this paper is to construct a model which guarantees the real determinants of the quark mass matrices assuming a non-invertible symmetry.
}

\makeatletter
\gdef\@fpheader{}
\makeatother

\begin{document}

\maketitle

\section{Introduction}
\label{sec:Intro}

The fundamental theory of particle physics may be CP invariant.
For example, string theory as well as a certain type of higher dimensional theory has CP symmetry \cite{Green:1987mn,Strominger:1985it}.
CP symmetry can be violated by moduli stabilization through compactifications.
However, CP violating vacua are hardly realized by simple moduli stabilization.
In particular, if the theory has other symmetries such as modular symmetries, CP invariant vacua are favorable.
(See e.g. Refs.~\cite{Kobayashi:2019uyt,Kobayashi:2020uaj,Ishiguro:2020nuf,Novichkov:2022wvg,Knapp-Perez:2023nty,Higaki:2024pql}.)
That leads to CP invariant four dimensional low energy effective field theory at the compactification scale, 
where the QCD phase $\theta$ is vanishing and Yukawa couplings as well as other couplings are real.

It is very interesting if our universe is indeed CP invariant in the fundamental theory at high energies. It means that the CP invariance must be spontaneously broken down at an intermediate energy scale, since we know the CP is violated in various meson decays at low energies. This fact stimulates us to consider the baryon-number asymmetry in the present universe is dynamically created in the early universe \cite{Yoshimura:1978ex}. We may understand the small baryon asymmetry in the present universe.

Furthermore, the above hypothesis would open a new window to solve the strong CP problem without the QCD axion \cite{Liang:2024wbb}. The experimental constraint on the physical vacuum-angle $\bar \theta \le10^{-10}$ \cite{Abel:2020pzs} requires a strict constraint on the mass matrices for the quarks such as arg$[\det M_uM_d]\le 10^{-10}$  \cite{Abel:2020pzs}, where $\bar \theta = \theta + {\rm arg}[\det M_uM_d]$. 
If we have some mechanism to guarantee real determinants of the quark mass matrices, the vanishing physical vacuum angle $\bar \theta$ indicates the CP invariance  $\theta=0$ at the fundamental level. Thus,  the small ${\bar \theta}$ is technically natural, since we have an enhanced CP invariance in the limit of the vanishing $\bar \theta =0$. In fact, it was proved that the vacuum angle is never renormalized up to the four-loop level \cite{Ellis:1978hq}.

We need a proper form of quark mass matrices to realize arg$[\det M_uM_d]\le 10^{-10}$.
We consider, in this paper, a non-invertible symmetry to control the quark mass matrices to satisfy the above constraint in the framework of the standard model (SM).
In particular, we use $\mathbb{Z}_2$ gauging of $\mathbb{Z}_N$ symmetries \cite{Kobayashi:2024yqq,Kobayashi:2024cvp,Funakoshi:2024uvy}.
That is a novel symmetry to lead interesting coupling selection rules.
One can realize realistic quark and lepton mass textures \cite{Kobayashi:2024cvp,Kobayashi:2025znw,Kobayashi:2025ldi,Kobayashi:2025cwx,Nomura:2025sod,Chen:2025awz,Okada:2025kfm,Jangid:2025krp} including axion-less solutions to the strong CP problem \cite{Liang:2025dkm,Kobayashi:2025thd}.\footnote{See, e.g. Refs.~\cite{Gomes:2023ahz,Schafer-Nameki:2023jdn,Bhardwaj:2023kri,Shao:2023gho}, for reviews on non-invertible symmetries, and Refs.~\cite{Choi:2022jqy,Cordova:2022fhg,Cordova:2022ieu,Cordova:2024ypu,Delgado:2024pcv,Choi:2025vxr,Suzuki:2025oov,Kobayashi:2025lar,Suzuki:2025bxg}, for other applications of non-invertible symmetries in particle physics.}

By using a non-invertible symmetry, we propose a simple model for the axion-less solution of the strong CP problem. 
To simplify the charge assignment of the fermions, we take the representations of the SM fermions in the $SU(5)$ grand unification. We will not, however, unify all of the gauge interactions in the SM and we will not consider full GUT multiplets for scalar bosons.  A consistent formulation of this framework is discussed in Refs.~\cite{Ibe:2019ifm,Hotta:1995cd} and we call it the GUT-inspired SM. In addition to the SM gauge groups, we introduce the $U(1)_{B-L}$ gauge symmetry to explain naturally the seesaw mechanism to generate the observed small neutrino masses \cite{Minkowski:1977sc, Yanagida:1979as,Yanagida:1979gs,Gell-Mann:1979vob}.

This paper is organized as follows.
In section \ref{sec:selection rules}, we briefly review on 
$\mathbb{Z}_2$ gauging of $\mathbb{Z}_N$ symmetries.
In section \ref{sec:model}, we show our model.
In section \ref{sec:pheno}, we study phenomenological aspects of our model.
Section \ref{sec:con} is devoted to our conclusion.

\section{Non-invertible selection rules}
\label{sec:selection rules}

In this section, we review the non-invertible selection rules for matter fields. 
It was known that selection rules of twisted and/or untwisted strings in heterotic string theory on Abelian and non-Abelian orbifolds become non-invertible. 
Since they are labeled by conjugacy groups of the space group action~\cite{Dixon:1985jw,Dixon:1986jc,Hamidi:1986vh,Dixon:1986qv}, the selection rules are not understood in a group-like manner~\cite{Kobayashi:1990mc,Kobayashi:1991rp,Kobayashi:1995py,Kobayashi:2025ocp}. 
Similar phenomenon can be seen in the D-brane system \cite{Kobayashi:2024yqq,Funakoshi:2024uvy}. In the framework of type IIB string theory on toroidal orbifolds with magnetic fluxes, the fusion rule of momentum operators on orbifolds has a non-invertible structure, e.g., Fibonacci and Ising rules, depending on the value of magnetic fluxes~\cite{Kobayashi:2024yqq}. 

Following Refs.~\cite{Kobayashi:2024yqq,Kobayashi:2024cvp}, we introduce the non-invertible selection rule for matter fields. 
First, let us start from $\mathbb{Z}_N$ symmetry under which matter fields $\phi_i$ transform as
\begin{align}
    \phi_i \rightarrow g^{k_i} \phi_i.
\end{align}
Here, $g=e^{2\pi i/N}$ denotes the generator of $\mathbb{Z}_N$ and $k_i$ corresponds to $\mathbb{Z}_N$ charges. Note that the selection rule of two matter fields $\phi_i$ and $\phi_j$ obeys the group-like form $g^{k_i}g^{k_j}=g^{k_i+k_j}$.

Next, we focus on the automorphism of $\mathbb{Z}_N$, i.e., $\mathbb{Z}_2$,
\begin{align}
    r^2 = e,\quad
    rg^kr^{-1}=g^{-k},
\end{align}
where $r$ and $e$ respectively denote the $\mathbb{Z}_2$ generator and the identity. 
Among the conjugacy classes of $D_N\cong \mathbb{Z}_N\rtimes \mathbb{Z}_2$, we pick up the $\mathbb{Z}_2$ invariant one, i.e.,
\begin{align}
    [g^{(k)}] = \{r^n g^k r^{-n}| n=0,1 \}=\{g^k, g^{-k}\},
\end{align}
obeying the following multiplication rule:
\begin{align}
    [g^{(k_1)}]\otimes [g^{(k_2)}] = [g^{(k_1+k_2)}] + [g^{(k_1-k_2)}].
\end{align}
This selection rule is different from the group-like one, and in what follows, we call $\tilde{\mathbb{Z}}_N$ for a $\mathbb{Z}_2$ gauging of $\mathbb{Z}_N\rtimes \mathbb{Z}_2$. 
Note that the matter fields labeled by the above class $[g^{(k)}]$ correspond to the $\mathbb{Z}_2$ invariant mode: 
\begin{align}
    \Phi_j= \phi_j + \phi_{N-j}.
\end{align}
This $\mathbb{Z}_2$ symmetry is nothing but a geometric $\mathbb{Z}_2$ twist in the context of type IIB magnetized D-brane models on $T^2/\mathbb{Z}_2$.

When $N=4$ and 5, there are three independent classes.
In the following discussion, we analyze the case with $N=5$, i.e., $\tilde{\mathbb{Z}}_5$, where there exist three classes $\{[g^0], [g^1], [g^2]\}$ obeying
\begin{align}
    [g^0]\otimes [g^0] &= [g^0]\,,\qquad \qquad \qquad\qquad\qquad\,
    [g^0]\otimes [g^1] = [g^1]\otimes [g^0] = [g^1]\,,\nonumber\\
    [g^0]\otimes [g^2] &= [g^2]\otimes [g^0] = [g^2]\,,\qquad\qquad\,\,\,\,
    [g^1]\otimes [g^1] = [g^0] + [g^2]\,,\nonumber\\
    [g^1]\otimes [g^2] &= [g^2]\otimes [g^1] = [g^1] + [g^2]\,,\qquad
    [g^2]\otimes [g^2] = [g^0] + [g^1]\,.
\end{align}
These three independent classes can be assigned to three generations of fermions.
For example, when three generations of all left-handed fermion fields correspond to these three classes, $\{[g^0], [g^1], [g^2]\}$, one can realize the Yukawa matrix,
\begin{align}
\label{eq:Y5-1}
    \begin{pmatrix}
        \checkmark & 0 & 0 \\
        0 & \checkmark & 0 \\
        0 & 0 & \checkmark 
    \end{pmatrix},
\end{align}
for the Higgs field corresponding to $[g^0]$, 
\begin{align}
\label{eq:Y5-2}
    \begin{pmatrix}
         0 & \checkmark & 0 \\
        \checkmark & 0  & \checkmark\\
        0 & \checkmark & \checkmark 
    \end{pmatrix},
\end{align}
for the Higgs field with $[g^1]$, 
\begin{align}
\label{eq:Y5-3}
    \begin{pmatrix}
         0 & 0 & \checkmark \\
        0 & \checkmark   & \checkmark\\
        \checkmark & \checkmark & 0 
    \end{pmatrix},
\end{align}
for the Higgs field with $[g^2]$,
where the check symbol $\checkmark$ denotes allowed coupling coefficient.
If we combine the first one and the third one, we can obtain the Yukawa matrix, 
where only the $(2,2)$ entry is non-vanishing, and the others are vanishing.

\section{Model}
\label{sec:model}

In this section, we first construct a GUT-inspired SM model with a non-invertible symmetry for providing an axion-less solution to the strong CP problem. The model is very similar to the model in Ref.~\cite{Kobayashi:2025thd}.
Charges and class assignments of fermion and scalar fields are shown in Tables 
\ref{tab:charge} and \ref{tab:charge-scalar}.

\begin{table}[H]
    \centering
    \caption{Charges and class assignments of matter fields under $SU(5) \times U(1)_{B-L}$ and  $\tilde{\mathbb{Z}}_5^{(1)}\times \tilde{\mathbb{Z}}_5^{(2)}$  symmetries.}
    \label{tab:charge}
    \begin{tabular}{|c|c|c|c|}
    \hline
         & $T$ & $\overline{F}$ & $\bar N$ \\\hline
         $SU(5) \times U(1)_{B-L}$ & $({\bf 10},-1)$
         & $(\bar {\bf 5},3)$ & $({\bf 1},-5) $ \\ \hline
         $\tilde{\mathbb{Z}}_5^{(1)}$ & ($[g^0]$, $[g^1]$, $[g^2]$) & ($[g^0]$, $[g^1]$, $[g^2]$) & ($[g^0]$, $[g^1]$, $[g^2]$)  \\\hline
         $\tilde{\mathbb{Z}}_5^{(2)}$ & ($[g^0]$, $[g^1]$, $[g^2]$) & ($[g^0]$, $[g^1]$, $[g^2]$) & ($[g^0]$, $[g^1]$, $[g^2]$)  \\\hline
    \end{tabular}
\end{table}

\begin{table}[H]
    \centering
    \caption{Charges and class assignments of scalar fields under $SU(5) \times U(1)_{B-L}$ and $\tilde{\mathbb{Z}}_5^{(1)}\times \tilde{\mathbb{Z}}_5^{(2)}$  symmetries.}
    \label{tab:charge-scalar}
    \begin{tabular}{|c|c|c|c|c|}
    \hline
        &  $H_1$ & $H_2$ & $\Phi$  &$\eta$ \\\hline
         $SU(5) \times U(1)_{B-L}$ & $(\bar {\bf 5},-2)$ & $(\bar {\bf 5},-2)$ + ({\bf 45},-2) & $({\bf 1},10)$  &$({\bf 1},0)$  \\ \hline
         $\tilde{\mathbb{Z}}_5^{(1)}$ &  $[g^1]$ & $[g^2]$ & $[g^0]$ & $[g^2]$ \\\hline
         $\tilde{\mathbb{Z}}_5^{(2)}$  &  $[g^1]$ &$[g^0]$ & $[g^0]$ & $[g^1]$ \\\hline
    \end{tabular}
\end{table}

We introduce the SM Higgs boson called as $H_1$. 
Then, we have the following Yukawa couplings:
\begin{align}
    f_u TTH_1^\dagger 
+ f_dT\overline{F}H_1   
  ,
    \label{eq:SU5}
\end{align}
with
\begin{align}
    f_{u,d} =
    \begin{pmatrix}
        0 & \checkmark & 0\\
        \checkmark & 0 & \checkmark\\
        0 & \checkmark & \checkmark
    \end{pmatrix}
    \, .
\end{align}
We can confirm that both mass matrices, $M_u$ and $M_d$ are 4-zero textures  as,
\begin{align}
\label{eq:texture_before}
    {M^0}_{u,d}=\begin{pmatrix}
        0 & a_{u.d} & 0 \\
        a'_{u.d} & 0 & c_{u.d} \\
        0 & c'_{u.d} & d_{u.d}
    \end{pmatrix},
\end{align}
where $a_{u.d},a'_{u.d},c_{u.d},c'_{u.d}$ and $d_{u.d}$ are  all real constants.

We now introduce a complex scalar boson $\eta$ whose vacuum-expectation value (vev), $\langle\eta\rangle$, is complex which induces the spontaneous CP violation. This boson is a singlet under all the gauge groups, but it has non-trivial charges of the non-invertible symmetries as shown in Table \ref{tab:charge-scalar}. The possible renormalizable potential for it is given by
\begin{align}
\label{eq:Veta}
    V_\eta = & -\mu \eta \eta^\dagger-\mu'(\eta^2+\eta^{\dagger 2})
    +\xi (\eta \eta^\dagger)^2+\xi'(\eta^4 + \eta^{\dagger 4})+ \xi'_{33}(\eta^2 + \eta^{\dagger 2})\eta\eta^\dagger .
\end{align}
We see that the vev $\langle\eta\rangle$ has a complex phase and its absolute value is basically given by the scale of its mass. In this paper, we consider the vev $|\langle\eta\rangle| =O(10^{10}-10^{12})$ GeV and consider the inflation scale $H_{\rm inf}$ to be smaller than $|\langle\eta\rangle |$ to avoid the domain wall formation. (If we introduce a higher dimensional operator such as $\eta^5 + \eta^{\dagger 5}$ we do not have the domain wall problem.)

We show a mechanism which induces a coupling between the $\eta$ boson and the SM fermion proposed in Ref.~\cite{Liang:2024wbb}. We first introduce a heavy Higgs boson $H_2$ and couplings with $H_1$ and $\eta$ given by
\begin{align}
\label{eq:VHeta}
 V_{H-\eta}=\mu_{12}(H_1^\dagger H_2 \langle \eta \rangle + \mathrm{h.c.}) + \mu_{21}(H_1^\dagger H_2 \langle \eta ^\dagger \rangle + \mathrm{h.c.}) ,
\end{align}
where $\mu_{12}$ and $\mu_{21}$ are dimension one real parameters which we will take of the order of $|\langle\eta\rangle|$. The heavy Higgs $H_2$ has Yukawa couplings with the SM fermions, $T$ and ${\bar F}$, and its non-invertible charges are chosen so that its Yukawa couplings are localized \cite{Liang:2024wbb} in both matrices $M_{u,d}$ through the following term:
\begin{align}
    f_u'TTH_2^\dagger  +f_d^\prime T\overline{F}H_2
  .
    \label{eq:SU5-2}
\end{align}
Then, we have the following Yukawa terms:
\begin{align}
    f_{u,d}^\prime =
    \begin{pmatrix}
        0 & 0 & 0\\
        0 & \checkmark & 0\\
        0 & 0 & 0
    \end{pmatrix}
    \,,
\end{align}
where $f'_{u,d}$ are real. 
As a consequence, we have an effective Yukawa coupling of the SM Higgs $H_1$ after the integration of the heavy Higgs $H_2$, which are localized in (2,2) element as
\begin{align}
\label{eq:texture}
    M'_{u,d}=\begin{pmatrix}
        0 & a_{u,\,d} & 0 \\
        a'_{u,\, d} & b_{u,d}e^{i\phi_{u,\,d}} & c_{u,\,d} \\
        0 & c'_{u,\, d} & d_{u,\, d}
    \end{pmatrix},
\end{align}
where $a_{u.d},a'_{u.d},b_{u.d},c_{u.d},c'_{u.d}$ and $d_{u.d}$ are real. 
We can write  
\begin{align}
   & b_ue^{i\phi_u}=\frac{f'_u}{M_2^2}\,
    \left( \mu_{12}\langle H_1^\dagger\rangle \langle \eta\rangle
    +\mu_{21}\langle H_1^\dagger\rangle \langle \eta^\dagger\rangle\, \right), \notag \\ 
  &  b_de^{i\phi_d} =\frac{f'_d}{M_2^2}\, 
    \left( \mu_{12}\langle H_1\rangle \langle \eta^\dagger\rangle
    +\mu_{21}\langle H_1\rangle \langle \eta\rangle \right), 
\end{align}
where $b_{u,d}$ are absolute values of the right hand sides and $e^{i\phi_{u,d}}$ are their phases.
Note that $\langle H_1 \rangle$ is real, while $\langle \eta \rangle$ is complex.
Remarkably, we find $\phi_d=-\phi_u$.
We denote $\phi=\phi_d=-\phi_u$, where the phase $\phi$ is non-vanishing unless $\mu_{12}=\mu_{21}$. 
Here, $M_2$ is the mass of the heavy Higgs $H_2$ and we consider it to be of the order of $|\langle\eta\rangle|$.

A crucial point is that both matrices have complex phases and we can explain the CP violation in the CKM matrix. Furthermore, the determinants of both matrices are real and hence we can solve the strong CP problem as pointed out in Refs.~\cite{Liang:2024wbb,Liang:2025dkm,Kobayashi:2025thd}.

We ignore, throughout this paper, the cut-off suppressed higher order terms in the Lagrangian, since such terms are strongly dependent on the dynamics of the physics at the cut-off scale and also of the nature of the $\eta$.

We add the $U(1)_{B-L}$ gauge symmetry. This symmetry naturally requires three right-handed neutrinos $N_{Ri}~i=1-3$ to cancel the gauge anomalies. Here, the $N_R$ is defined as a charge conjugate state of the original ${\bar N}$. Its spontaneous breaking generates heavy Majorana masses for the $N_{Ri}$ which produce the observed small neutrino masses through the seesaw mechanism. We introduce a new scalar boson $\Phi$ whose vev gives the Majorana masses via the Yukawa coupling $\Phi^\dagger N_{Ri}N_{Ri}$. The quantum numbers of the $\Phi$ and the $N_{Ri}$ are shown in Tables \ref{tab:charge} and \ref{tab:charge-scalar}. The mass matrix for the $N_{Ri}$ is a real diagonal matrix
by our assignment of classes shown in Tables \ref{tab:charge} and \ref{tab:charge-scalar}. Notice that the CP-violating scalar $\eta$ never couples to $N_{Ri}N_{Rj}$, since it does not have the $U(1)_{B-L}$ charge.

\section{Phenomenology}
\label{sec:pheno}

We are ready to discuss the phenomenology of our model.

\subsection{The CKM matrix}

The mass matrices $M'_{u,d}$ for the up-type and the down-type quarks are given by Eq.~\eqref{eq:texture}.
As the usual convention, we define mass matrices 
as $\bar u_L M_u u_R$ and $\bar d_L M_d d_R$.  Thus, we have $M_{u,d}= (M'_{u,d})^*$, which are 
\begin{align}
\label{eq:quark}
    M_{d}=\begin{pmatrix}
        0 & a_{d} & 0 \\
        a'_{d} & b_{d}e^{-i\phi} & c_{d} \\
        0 & c'_{d} & d_{d}
    \end{pmatrix}, \qquad
    M_{u}=\begin{pmatrix}
        0 & a_{u} & 0 \\
        a_{u} & b_{u}e^{i\phi} & c_{u} \\
        0 & c_{u} & d_{u}
    \end{pmatrix}.
\end{align}
 We also consider the quark sector and the lepton sector together as $SU(5)$ multiplets.
 That is, we consider three $SU(5)$ multiplets $\bar F= \bf 5^*$ and $T=\bf 10$.
Then, the up-type quark mass matrix is symmetric,
that is $a_u=a'_u$ and  $c_u=c'_u$ in Eq.~\eqref{eq:texture}.
From the data on the masses and the observed CKM matrix elements
\cite{Antusch:2013jca}, we determine all 
parameters in the above matrices.

The numerical values of parameters in Eq.~\eqref{eq:texture}
are obtained as shown  in Table \ref{tab:parameters}.

\begin{table}[hbtp]
\caption{The allowed range of parameters in Eq.~\eqref{eq:texture},
where $2\,\sigma$ error bars of the quark masses and the CKM angles and  CP phase are taken in Refs.~\cite{ParticleDataGroup:2024cfk,Antusch:2013jca}.
}
\hskip 1.5 cm
\begin{tabular}{|c|c|c|c|c|}
\hline 
	\rule[14pt]{0pt}{3pt}  
 $a_d/d_d \times 10^{2}$ & $a'_d/d_d\times 10^{2}$   & $b_d/d_d\times 10^{2}$  &  $c_d/d_d\times 10^{2} $ & $c'_d/d_d$ \\
\hline 
	\rule[14pt]{0pt}{3pt}  
 $0.71 - 0.84$ &$0.48 - 0.58$& $5.1- 6.3$ 
 &$5.0 - 6.1$  & $0.89 - 1.13$\\
\hline
\end{tabular}

\hskip 1.5 cm
\begin{tabular}{|c|c|c|c|}
\hline 
	\rule[14pt]{0pt}{3pt}  
 $a_u/d_u \times 10^{4}$ &$b_u/d_u \times 10^{3}$ &$c_u/d_u \times 10^{2}$ &$\phi$ $^{[\circ]}$\\
\hline 
	\rule[14pt]{0pt}{3pt}  
$0.78-1.04$ & $2.4-3.2$& $1.2-1.6$ &$32 - 42$\\
\hline
\end{tabular}
\label{tab:parameters}
\end{table}
The number of parameters in the up- and down-type quark mass matrices is eleven, but the observable parameters are the six quark masses and the four CKM matrix elements. This is a reason why the determined parameters have relatively large error bars.

\subsection{The charged lepton mass matrix}
We consider the charged lepton mass matrix.
Here, we have assumed that the Higgs bosons $H_1$ and $H_2$ belong to $\bf 5^*$ of the $SU(5)_{\rm GUT}$. We see immediately the wrong mass relation, that is, $m_e=m_d$ and $m_\mu=m_s$. Thus, we consider that the heavy Higgs boson $H_2$ is not purely the $\bf 5^*$ but a mixture with $\bf 45$ of the $SU(5)$. Then we have one real free parameter $k_e$ and the mass matrix is given by
\begin{align}
M_e=
\begin{pmatrix}
0 & a' & 0 \\
a & ~k_e\, b \ e^{-i\phi}& c'\\
0 & c & d
\end{pmatrix}\ , \quad
\label{Me}
\end{align}
where
we multiply the (2,\,2) element by a factor $k_e$. 
The correct mass ratios of the electron and the muon to the tau
are obtained by setting  $k_e=3$. It is surprising that the two independent precise mass ratios are explained by choosing one free parameter $k_e$. See Refs.\,\cite{Tanimoto:2024nwm,Tanimoto:2025fnj} for details.

\subsection{The CP violation in the neutrino oscillations}

As shown above, the mass matrix for the heavy Majorana right-handed neutrinos $M_{N_R}$ is a real diagonal matrix, that is, $M_{N_R}={\rm diag}(M_1,M_2,M_3)$. And the Dirac type mass matrix $M_D$ is given by 
\begin{align}
 M_D=
\begin{pmatrix}
0 & a'_\nu & 0 \\
a_\nu & b_\nu \ {e^{i\phi}}& c'_\nu\\
0 & c_\nu & d_\nu
\end{pmatrix}\,.
\label{MD}
\end{align}
The light neutrino mass matrix $M_\nu$ is given (via the seesaw mechanism) by:
\begin{align}
M_\nu=
\begin{pmatrix}
A'^2 & A'B  \,{e^{i\phi}}& A' C \\
 A'B  \ {e^{i\phi}}&  A^2+ C'^2+B^2\, {e^{2i\phi} } &B C \,{e^{i\phi}}+C' D\\
  A' C & B C \,{e^{i\phi}}+C' D\ & D^2+C^2
\end{pmatrix}\,,
\label{Mnu}
\end{align}
where the parameters   are defined
by $A= a_\nu/\sqrt{M_1}$, $A'= a'_\nu/\sqrt{M_2}$, $B= b_\nu/\sqrt{M_2}$,  $C= c_\nu/\sqrt{M_2}$, $C'= c'_\nu/\sqrt{M_3}$  
and $D= d_\nu/\sqrt{M_3}$. 

\subsection{Predictions of the CP phase  and the neutrino-less double beta decay}

Since the charged lepton mass matrix and neutrino mass matrix are given in Eqs.~(4.2) and (4.3),
we can discuss the CP violation of neutrinos and 
 the effective mass 
 in  the neutrino-less double beta decays $m_{\beta\beta}$
 by inputting data of leptons in Ref.~\cite{Esteban:2020cvm}.
Then, we  obtain the allowed ranges of parameters   by scanning them in the real space with
 $|D| \geq |A|, \, |A'|, \,|B|, \, |C|, \, |C'|$ in order to realize  NH of neutrino masses.

\begin{wrapfigure}{r}{8cm}
\vspace{0mm}
\includegraphics[width=7.5cm]{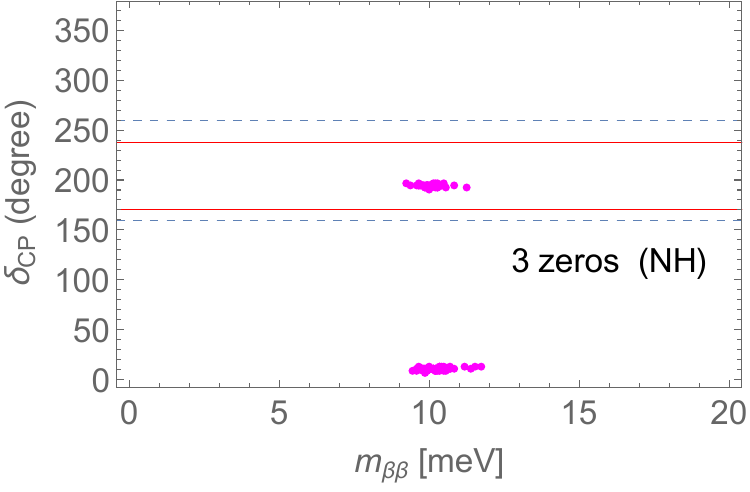}
\vskip -0.3 cm
\caption{The predicted $\delta_{\rm CP}$ versus  $m_{\beta\beta}$
for NH.
The region between the  horizontal red (blue dashed) lines 
denotes $1\,(2)\sigma$ allowed one of $\delta_{\rm CP}$ in NuFIT 6.0 
\cite{Esteban:2020cvm}.}
\label{mee-CP}
\end{wrapfigure}

Then, we  predict the CP-violating phase $\delta_{\rm CP}$ in the neutrino oscillation and the mass parameter for the neutrino-less double beta decay $m_{\beta\beta}$ 
by using the phase $\phi$ determined from the CKM matrix. 
The predicted  $\delta_{\rm CP}$ versus   $m_{\beta\beta}$  are shown 
in Fig.\ref{mee-CP}.
Since the real parameters are allowed in both plus and minus signs,
$\delta_{\rm CP}$ is allowed in two regions $180^\circ$ apart.
Since one region is in the middle of the experimental allowed region 
with $1\,\sigma$, we exclude another region.

Thus, the predicted range of $\delta_{\rm CP}$ is:
\begin{equation}
   \delta_{\rm CP}\simeq 192^\circ - 197^\circ \,.
   \label{CP}
\end{equation}
It is remarkable that the prediction is within $1\,\sigma$ range of 
in NuFIT 6.0 \cite{Esteban:2020cvm}.

The predicted  $m_{\beta\beta}$  is: 
\begin{equation}
    m_{\beta\beta} \simeq 9-11\  [{\rm meV}]\,,
     \label{mee}
\end{equation}
which will be tested in  future experiments.
It is noted that $m_{\beta\beta}$ is almost proportional to  $\sum  m_i$.
The predicted range of $\sum  m_i$ is $69-74$\,meV.

\vskip 0.5 cm
 These predictions may be compared with previous ones in Ref.\,\cite{Tanimoto:2024nwm}, where the down-type quark mass matrix is the same as the one in Eq.\,\eqref{eq:quark}, but the up-type quark mass matrix is diagonal:

\begin{equation}
   \delta_{\rm CP}\simeq 192^\circ - 195^\circ \,, \qquad   m_{\beta\beta} \simeq 8-11\  [{\rm meV}]\,.
   \label{previous}
\end{equation}
Our predictions of the present model in Eqs.\,\eqref{CP} and \eqref{mee} are a little bit different from previous ones of Eq.\,\eqref{previous}. This is because the texture of the up-type quark mass matrix there is real diagonal, and the CKM matrix is given only by the down-type quark mass matrix.

We present the magnitude of neutrino parameters 
in Table \ref{tab:neu}, where the magnitudes of parameters
are almost of the same order except for $C'$.

\begin{table}[hbtp]
\begin{center}
\caption{The allowed regions of parameters in  $M_\nu$.}
\begin{tabular}{|c|c|c|c|c|}
\hline 
	\rule[14pt]{0pt}{3pt}  
 $|A/D|$& $|A'/D|$ & $|B/D|$ &  $|C/D|$& $|C'/D|$\\
\hline 
	\rule[14pt]{0pt}{3pt}  
 $0.474-0.498$ & $0.464-0.495$ & $0.093-0.104$& $0.290-0.335$ 
 & $0-0.053$\\
\hline
\end{tabular}
\label{tab:neu}
\end{center}
\end{table}

\subsection{The universe's baryon asymmetry}
\label{Baryon}

The decays of the heavy right-handed neutrinos (RHNs), $N_{Ri}$, generate the lepton asymmetry which is converted to the baryon-number asymmetry in the Universe~\cite{Fukugita:1986hr}. The dominant asymmetry is generated by the decay of the lightest RHN and thus we have to identify the lightest  RHN to calculate the baryon asymmetry. We show that the CP phase relevant to the generation of the baryon asymmetry is directly related to the low-energy CP phase in the CKM and PMNS mass matrices.

Let us assume that the $N_{R1}$ is the lightest RHN. The induced lepton asymmetry $\epsilon_1$ is estimated as \cite{Plumacher:1996kc,Hamaguchi:2002vc}
\begin{equation}
    \epsilon_1
    =\frac{\Gamma(N_{R1}\to L H)-\Gamma(N_{R1}\to \bar L \bar H)}
    {\Gamma(N_{R1}\to L H)+\Gamma(N_{R1}\to \bar L \bar H)}\simeq -\frac{1}{8\pi} \sum _{j\not= 1}^3
    \frac{{\rm Im}[(Y_D^\dagger Y_D)^2_{1j}]}{(Y_D^\dagger Y_D)_{11}}
    F^{V+S}\left(\frac{M_j^2}{M_1^2}\right )\,,
    \label{asymmetry}
\end{equation}
where  $Y_D=M_D/v$, $v$ being the vacuum expectation value of the standard Higgs, and
\begin{equation}
    F^{V+S}(x)=\sqrt{x}\left [(x+1)\ln\left (\frac{x+1}{x}\right )-1+\frac{1}{x-1}\right ]
     \xrightarrow[x\rightarrow\infty]{} \frac32 \frac{1}{\sqrt{x}}\,.
\end{equation}
The baryon-number asymmetry is given by
\begin{equation}
    Y_B \simeq -\frac{28}{79}\, \kappa \,\frac{\epsilon_1}{g^*}\,,
\end{equation}
where $\kappa$ is a dilution factor which involves the integration of the full set of Boltzmann equations,   and $g*=106.75$ is taken in SM.

Since only the (2,\,2) element of $M_D$ has the complex phase $\phi$,   the non-vanishing complex phase appear in $(Y_D^\dagger Y_D)^2_{12}$ 
 as:
\begin{equation}
   {\rm Im}[(Y_D^\dagger Y_D)^2_{12}] = {\frac{1}{v^4}\, a_\nu^2\, b_\nu^2\,\sin 2\phi }\,,
\end{equation}
which is {positive} because of $\phi=32^{\circ}-42^{\circ}$ as seen in Table \ref{tab:parameters} and hence the $\epsilon_1$ is {negative}. 

Now, we see that our model predicts the right sign of the baryon-number asymmetry if the $N_{R1}$ is the lightest RHN.
It is emphasized that the CP-violating phase for creating the baryon asymmetry is exactly the same as the CP-violating phase $\phi$, which is determined from the observed CKM phase.

\section{Conclusions}
\label{sec:con}

We have constructed a three-zero texture model for the quark/lepton mass matrices imposing a non-invertible  $\tilde{\mathbb{Z}}_5^{(1)}\times \tilde{\mathbb{Z}}_5^{(2)}$ symmetry. The determinants of all quark mass matrices are real and hence the observed smallness of the physical $\bar \theta$ means we have a complete CP invariance at the fundamental level\footnote{The three zero textures are broken by quantum corrections. It happens at the two loop level as pointed out in Ref.~\cite{Liang:2025dkm}. However, owing to the small Yukawa coupling constants, it is sufficiently small to solve the strong CP problem.}.

Our model construction is based on a $SU(5)$-GUT inspired SM and the inclusion of the charged lepton and the neutrino sectors is almost straightforward. This simple model has been successful in explaining naturally the correct sign of the baryon-number asymmetry in the present universe. It also provides us with the predictions on the CP-violating phase $\delta_{\rm CP}$ and on a key parameter $m_{\beta \beta} $  as shown in Eqs.~\eqref{CP} and \eqref{mee}.

A hope is to test the above predictions in future experiments. However, we should notice here that the predictions in the neutrino sector depend on the charge assignment of the non-invertible symmetry for the right-handed neutrinos, $N_{Ri}$ and the Higgs boson $\Phi$. If the predictions on $\delta_{\rm CP}$ and on $m_{\beta \beta} $ in the present paper turn out to be excluded in future experiments, we will come back to the issue. 


\acknowledgments

This work was supported by JSPS KAKENHI Grant Numbers JP23K03375 (T.K.), JP25H01539 (H.O.), and JP24H02244 (T.T.Y.). T.~T.~Y.~was supported also by the Natural Science Foundation of China (NSFC) under Grant No.~12175134 as well as by World Premier International Research Center Initiative (WPI Initiative), MEXT, Japan.

\bibliography{references}{}
\bibliographystyle{JHEP}

\end{document}